\documentclass[runningheads,a4paper]{llncs}
\usepackage{makeidx}  
\usepackage{graphicx}
\begin{document}
\mainmatter              
\title{Simulation-based Validation of Smart Grids -- Status Quo and Future Research Trends}
\titlerunning{Simulation-based Validation of Smart Grids}  
\author{
	C.~Steinbrink\inst{1} \and
	S.~Lehnhoff\inst{1} \and
	S.~Rohjans\inst{2} \and
	T.~I.~Strasser\inst{3} \and				
	E.~Widl\inst{3} \and    
	C.~Moyo\inst{3} \and
	G.~Lauss\inst{3} \and
    F.~Lehfuss\inst{3} \and
	M.~Faschang\inst{3} \and 
	P.~Palensky\inst{4} \and
	A.~A.~van~der~Meer\inst{4} \and
	K.~Heussen\inst{5} \and
	O.~Gehrke\inst{5} \and
    E.~Guillo-Sansano\inst{6} \and
	M.~H.~Syed\inst{6} \and
	A.~Emhemed\inst{6} \and
	R.~Brandl\inst{7} \and
	V.~H.~Nguyen\inst{8} \and 
    A.~Khavari\inst{9} \and 
    Q.~T.~Tran\inst{10} \and
    P.~Kotsampopoulos\inst{11} \and
    N.~Hatziargyriou\inst{11} \and 
    N.~Akroud\inst{12} \and 
    E.~Rikos\inst{13} \and 
    M.~Z.~Degefa\inst{14}
} 
\authorrunning{C.~Steinbrink et al.} 
\institute{
	OFFIS e.V., Oldenburg, Germany \and
    HAW Hamburg University of Applied Sciences, Hamburg, Germany \and
	AIT Austrian Institute of Technology, Vienna, Austria \and
	Delft University of Technology, Delft, The Netherlands \and
	Technical University of Denmark, Lyngby, Denmark \and
	University of Strathclyde, Glasgow, United Kingdom \and
	Fraunhofer Inst. of Wind Energy and Energy System Technology, Kassel, Germany \and
	University Grenoble Alpes, G2Elab, Grenoble, France \and 
    European Distributed Energy Resources Lab. (DERlab) e.V., Kassel, Germany \and 
    Commissariat à l'énergie atomique et aux énergies alternatives, Chambery, France \and
    National Technical University of Athens, Athens, Greece \and 
    Ormazabal Corporate Technology, Bilbao, Spain \and 
    Centre for Renewable Energy Sources and Saving, Athens, Greece \and 
    SINTEF Energy Resarch, Trondheim, Norway \\[0.15cm]
	\email{cornelius.steinbrink@offis.de}
}
\maketitle              
%
%
\begin{abstract}
Smart grid systems are characterized by high complexity due to interactions between a traditional passive network and active power electronic components, coupled using communication links. Additionally, automation and information technology plays an important role in order to operate and optimize such cyber-physical energy systems with a high(er) penetration of fluctuating renewable generation and controllable loads. As a result of these developments the validation on the system level becomes much more important during the whole engineering and deployment process, today. In earlier development stages and for larger system configurations laboratory-based testing is not always an option. Due to recent developments, simulation-based approaches are now an appropriate tool to support the development, implementation, and roll-out of smart grid solutions. This paper discusses the current state of simulation-based approaches and outlines the necessary future research and development directions in the domain of power and energy systems. 
\end{abstract}
%
%
\section{Introduction}
\label{sec:introduction}
Due to the interactions between the traditional passive network and also of active power electronic components via dedicated communication networks, smart grids tend exhibit a  high degree of complexity \cite{Farhangi2010}. Sophisticated automation and information technology, corresponding control algorithms and data analytics methods are also of high importance for the reliable operation and optimization of such cyber-physical energy systems. This is so as to cope with a high(er) penetration of fluctuating renewable generation and controllable loads. As a result of these developments the validation on the system level, i.e., the testing of the integration and interaction of the connected components and algorithms, becomes today much more important during the whole engineering and deployment process \cite{Palensky2017,Strasser2017}. In earlier development stages and for larger system configurations laboratory-based testing is not always an option. Simulation in the domain of power systems is fundamental in order to understand system behaviour under normal but also in emergency situations. It also avoids costly and time-consuming real-world laboratory testing or field trials \cite{Strasser2017}. Due to recent developments, simulation-based approaches are an important tool in the development, implementation, and roll-out of smart grid solutions \cite{Podmore:2010}.

This paper provides a comprehensive discussion of the current state of simu-lation-based validation approaches in the domain of power and energy systems and addresses smart grid validation needs. It identifies shortcomings in today's practice and outlines the necessary future research and development steps. 

The rest of this paper is organized as follows: Section~\ref{sec:validation Challenges} briefly outlines main challenges in the validation of smart grid systems. A comprehensive overview of simulation-based smart grid development and validation approaches is given in Section~\ref{sec:co-simulation} followed by a discussion of future research needs and directions. Section~\ref{sec:conclusions} concludes the paper with the key findings.
%
%
\section{Validation Challenges}
\label{sec:validation Challenges}
Traditionally, the separate domains of power system and Information and Communication Technology (ICT)/automation have been analysed individually. In the context of the smart grid advancement, and for the first time, new requirements now demand simultaneous coverage of both domains in a comprehensive system-level validation. As already pointed out in the introduction, simulation-based approaches play a vital role in enabling this \cite{Palensky2017,Podmore:2010}.  

\begin{figure}[!htbp]
	\centering
	\includegraphics[width=0.95\columnwidth]{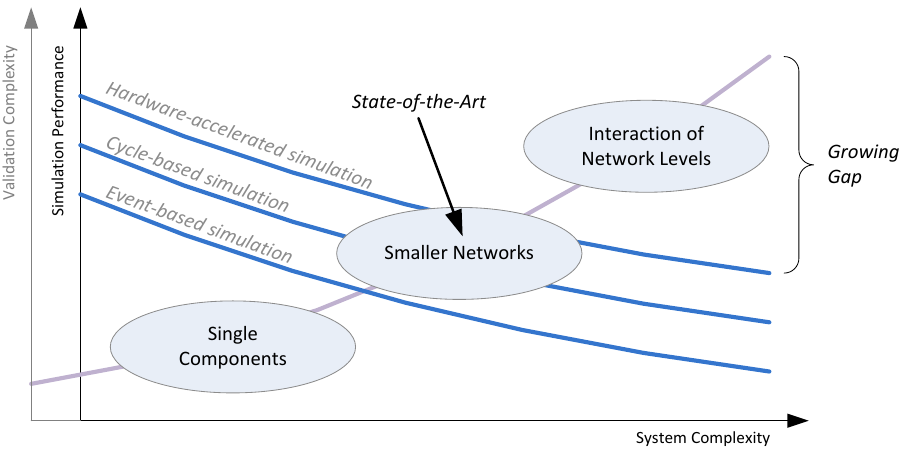}
	\caption{Growing gap between system complexity and performance of simulations}
	\label{fig:validation_complexity}
\end{figure}

Figure~\ref{fig:validation_complexity} shows the state-of-the-art in smart grid simulation and corresponding validation. The system complexity grows with the extension of the analysed network part. Smart grid systems will demand modelling of the interaction of different network levels but also the integrated analysis of ICT issues. 

A fast growing gap can be observed between the system complexity and the performance of known methods for system simulation (see Figure~\ref{fig:validation_complexity}). It still remains that for the hardware-accelerated simulation, a feature of real-time hardware-in-the-loop simulations, the interaction of network levels can only be analysed at very coarse granularity. Today’s tools still stem from the phase where single domains were in the focus. As soon as they get coupled with tools from other domains (ICT/automation, energy markets, customer behaviour, etc.), the performance is usually considerably reduced due to the necessary data exchange and format conversion.

In the following sections an overview of different simulation approaches is provided and future research needs and directions are derived in order to overcome the above mentioned gap. 
%
%
\section{Simulation-based Validation Approaches}
\label{sec:simulation-based_validation_approaches}
Simulation approaches used in the power and energy domain can generally be divided into the following three areas: \textit{(i)} multi-domain simulation, \textit{(ii)} cooperative simulation (also known as co-simulation), and \textit{(iii)} real-time simulation and hardware-in-the-loop. An overview of all these approaches is given below.
\subsection{Multi-Domain Simulation}
\label{sec:multi-domain_simulation}
Pure simulation tools are suitable in the early development stages for concept design and proof-of-concept validation. Purely analytical or numerical tools are suitable for concept analysis, but if there is no clear separation of the solution (e.g., an embedded control system) the simulation results cannot serve as a means of validation. The validation of embedded systems solutions will require the integration of heterogeneous simulation components as both the systems complexity and heterogeneity is increasing and specialized and validated simulation models are typically developed in a single domain. The simulation framework Ptolemy II \cite{davis1999ptolemy} offers a rigorous solution to this integration by focusing on the determinism of simulation outcome -- a feature particularly important for validation purposes. Another important multi-domain simulation has been developed with the Modelica language \cite{elmqvist1997introduction}. It is also based on first principles approaches in the declaration of models, strictly separating an object and interface-oriented approach to ``modelling'' from ``solving'', which is performed in a compilation step. Modelica is supported by both active open source development \cite{fritzson2011modelica} as well as commercial packages. The third example for a  purpose-built multi-domain simulator for smart grid algorithms is  IPSYS \cite{ipsys2004,ipsys2014} and it is meant for performance-assessment of hybrid power system control strategies. Multi-domain simulators are each built on generalized principles of interaction. 

Despite the powerful uses in performance assessment, system-level validation is questionable, as a multi-domain simulator typically requires the candidate system to be adapted or even reformulated to comply with the respective simulator architecture. 
\subsection{Co-Simulation}
\label{sec:co-simulation}
\noindent \textit{Overview and Distinction with other Simulation Types \\[-0.5em]}

\begin{figure}[!htbp]
	\centering
	\includegraphics[width=0.90\columnwidth]{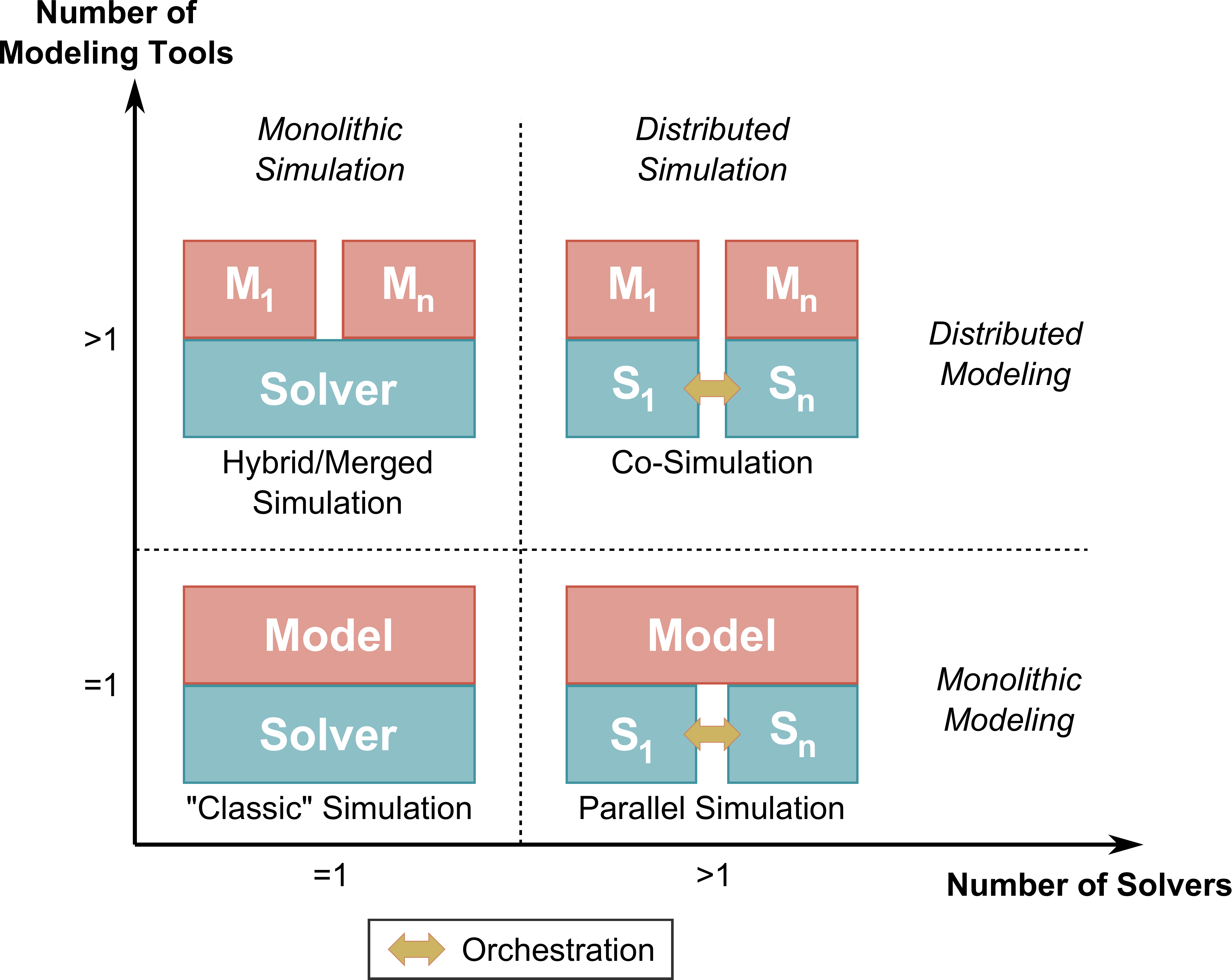}
	\caption{Distinction between co-simulation and other simulation types \cite{Geimer2006,schloegl2015}}
	\label{fig:sim_overview}
\end{figure}

\noindent Co-simulation is defined as the coordinated execution of two or more models that differ in their representation as well as in their runtime environment \cite{schloegl2015}. Representation in this context means the underlying modelling paradigm. For example, models may be represented as differential equation systems, discrete automata, etc. A runtime environment is a software system that solves model equations or generally allows the model execution. The models in a co-simulation system, therefore, have been developed as well as implemented independently. A number of simulation concepts are related to co-simulation, but differ slightly in their definitions \cite{Geimer2006}. Setups with combined development and implementation of all model systems are used for ``classic'' simulation. If models are developed jointly but are then separated into different runtime environments, one speaks of distributed or parallel simulation. Joint implementation and execution of models with different representation is sometimes called hybrid or merged simulation. The different types of simulation are depicted in Figure~\ref{fig:sim_overview}.

It is important to note that co-simulation implies the interaction of hardware and software components in some domains. In general, all components of a co-simulation setup may be either hardware or software. If hardware/software co-simulation is conducted for hardware testing, it is typically called ``Hardware-in-the-Loop'' (HIL) \cite{de2011}.

The major benefit of co-simulation is the separation of the modelling and simulation processes. Different researchers or even institutes may develop and implement simulation models representing different systems. Co-simulation users may then employ these models to analyse the dynamics of larger ``systems of systems''. In other words, co-simulation supports reuse of simulation models. Ideally, these models have been created by experts of the particular domain, are properly validated, and thus acknowledged. \\[-0.5em]

\noindent \textit{Generic vs. Specific Co-Simulation \\[-0.5em]}

\noindent The driving force behind co-simulation is the fact that the involved subdomains already utilise numerous established technologies and simulation tools. Independently, the interactions of each subdomain are not yet fully understood. Many co-simulation approaches feature manual coupling of a small number of tools ad-hoc. Mainly power system and communication network simulation frameworks are regarded \cite{lin2011,godfrey2010,georg2013,mets2011} since they are considered the major determining factors in smart grid dynamics.

\begin{figure}[!htbp]
	\centering
	\includegraphics[width=0.85\columnwidth]{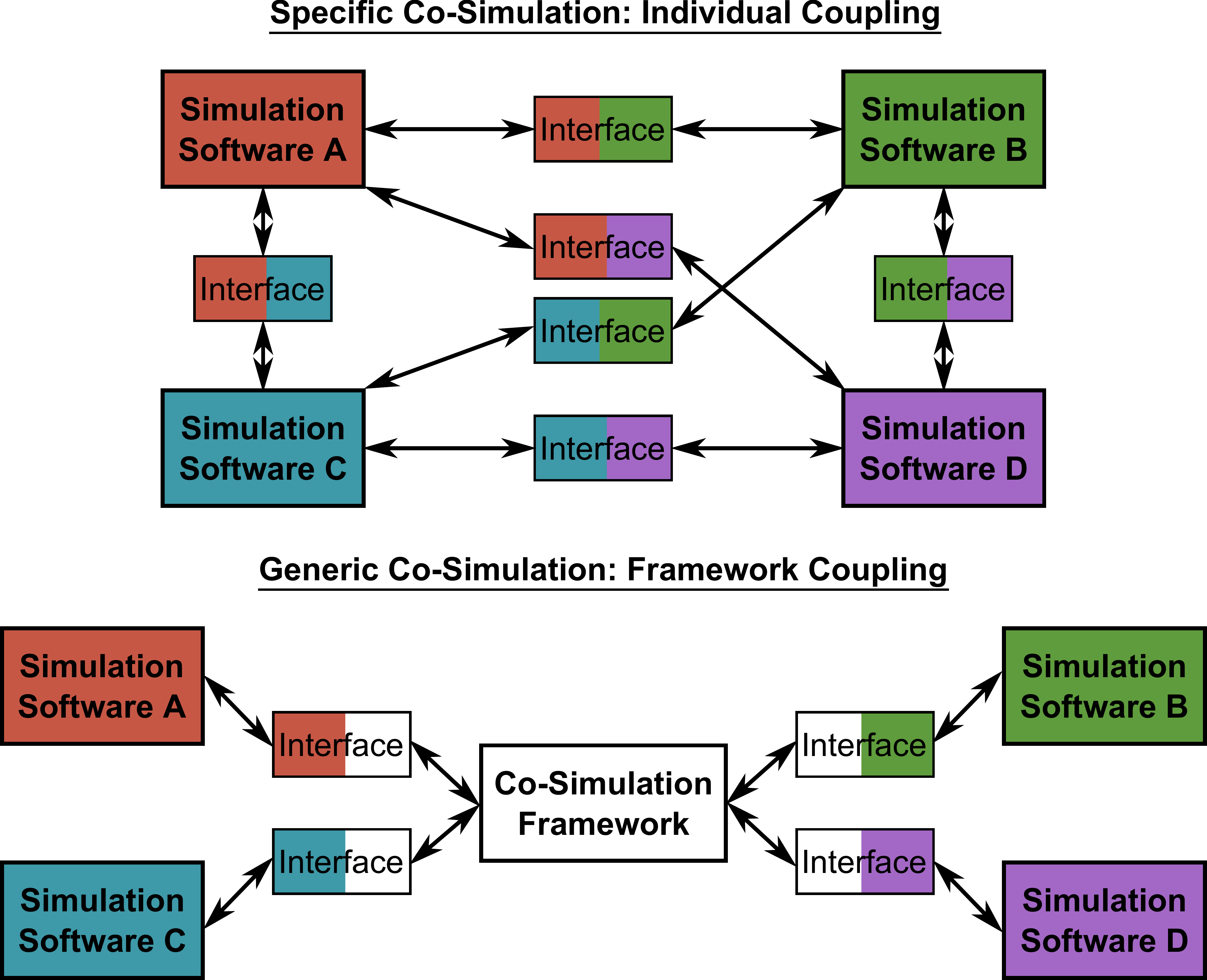}
	\caption{Difference between specific and generic co-simulation \cite{steinbrink2016}}
	\label{fig:sim_couple}
\end{figure}

Other co-simulation projects, however, support a more generic approach with a stronger inclusion of different system components. Generic co-simulation approaches involve a middleware that is responsible for data exchange and temporal synchronization of several  models. A software fulfilling these tasks is called a co-simulation framework. It typically provides a set of interfaces that may be implemented to establish a connection between the framework and a given model. A connected model can then indirectly exchange data with all other connected tools. The framework’s synchronization algorithm provides a common time frame for this exchange. Therefore, individual coupling between simulation tools is not necessary. This greatly reduces the likelihood of coupling errors and allows easy reuse of tools in various co-simulation studies. The difference between the manual ad-hoc coupling and the framework coupling is shown in Figure~\ref{fig:sim_couple}. \\[-0.5em]

\noindent \textit{Model Instantiation \\[-0.5em]}

\noindent An important concept in co-simulation is model instantiation. That means that several virtual objects are derived from the same model to simulate the behaviour of a number of similar systems. Such an object is called entity or instance. One example is that of several virtual Photovoltaic (PV) panels that are all derived from the same simulation model. The instantiation is important since each entity might receive different inputs from a spatial resolved solar irradiation model. In other words, instantiation is used in co-simulation to account for complex inputs and boundary conditions affecting relatively simple component models. Sometimes, however, the implementation of a simulation tool does not allow for multiple instantiation of it's models. These models are then called singletons. \\[-0.5em]

\noindent \textit{Interfaces for Simulator Abstraction \\[-0.5em]}

Next to software for co-simulation orchestration and execution, interfaces for simulator abstraction are an important concept in co-simulation research. The most popular standard in this context is the Functional Mockup Interface (FMI) \cite{blochwitz2011}. It has been defined in the industrially led MODELISAR project in coordination with the modelling language Modelica. Thus, FMI is supported by Modelica environments like Dymola \cite{bruck2002} or SimulationX \cite{noll2011}, but also by independent tools and languages like Matlab and Python. Some tool-specific co-simulation approaches employ FMI as well for interface descriptions in order to facilitate future extensibility of their setups \cite{bastian2011}. 
The FMI’s benefit is not just given by its formalism in model description. The standard allows users to make their models and simulation tools accessible in the form of so-called Functional Mock-up Units (FMU) that contain an FMI-based formalization as well as some form of representation of the tool in question. This representation depends on the type of FMI employed. The FMI standard is divided into two main parts: \textit{(i)} FMI for Model Exchange and \textit{(ii)} FMI for Co-Simulation. The essential difference is that FMUs of the former standard expect to be solved by a given master algorithm. FMUs of the latter standard, on the other hand, contain a solver so that the master algorithm is only required for the coordination of data exchange. In other words, models standardized with FMI for Co-Simulation are co-simulation-ready components while those standardized with FMI for Model Exchange are not. 

All in all, it can be said that co-simulation frameworks (popular tools are Mosaik \cite{rohjans2013,schutte2011}, Ptolemy II \cite{ptolemaeus2014}, and C2WT-TE \cite{neema2016}) and standards are subjects to active, ongoing research. Many tools and concepts already exist with varying foci, features, and degrees of usability and popularity. 
\subsection{Real-time Simulation and Hardware-in-the-Loop}
\label{sec:real-time_simulation_and_hil}
A simulation where a fixed time-step of the simulator, required for achieving a solution and performing I/O activities, is equal to the actual wall-clock time is commonly referred to as a real-time simulation. For validation purposes, the real-time simulation is commonly coupled with a Hardware-under-Test (HUT), adding the complexity of the hardware to the assessment procedure \cite{de2011}. This advanced testing method, HIL, allows for an extensive analysis of the HUT while under a simulated broad range of operating conditions, reducing the risk associated with performing tests on an actual network and at the same time reduces the cost and time required for performing validation of power system components and energy systems.

\begin{figure}[!htbp]
	\centering
	\includegraphics[width=0.90\columnwidth, trim={0 1.5cm 0.5cm 0.9cm},clip]{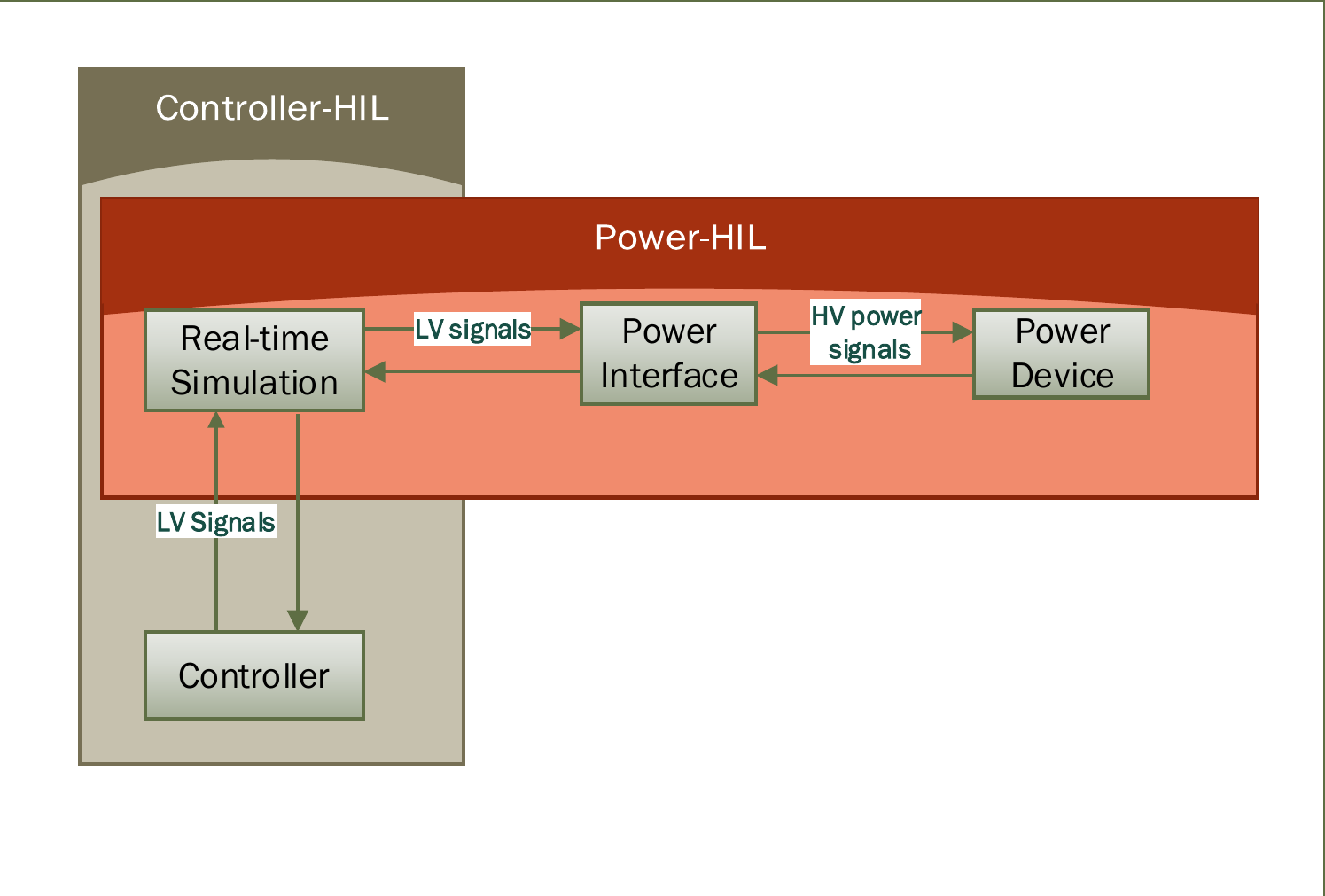}
	\caption{Overview of the CHIL and PHIL simulation concepts}
	\label{fig:HIL}
\end{figure}

Depending on the characteristics of the HUT and the properties of interface between HUT and the real-time simulation, HIL can be classified as controller or power HIL (shown in Figure~\ref{fig:HIL}).

\begin{itemize} 
	\item \textit{Controller-HIL (CHIL)}, the HUT exchanges low voltage signals (+/-10V) with the real-time simulation. The HUT in CHIL is typically a controller device, although real-time simulations coupled to other devices such as relays, PMU or monitoring components are usually classified as CHIL. This devices are validated in a closed-loop environment under different dynamic and fault conditions, therefore enhancing the validation of control and protection systems for power systems and energy components \cite{Steurer:2007}.
    \item \textit{Power-HIL (PHIL)}, the exchanged signals between simulation and HUT are of high power and therefore a dedicated power interface for amplifying the exchanged signals and injecting or absorbing power is required for coupling high power HUT with software simulation. Different approaches can be selected for exchanging the signals, known as interface algorithms, depending on the application and the compromise between stability and accuracy of the simulation \cite{Ren2008}. The addition of the power interface introduces inaccuracies into the process as the power interface cannot achieve unity gain with infinite bandwidth and zero time delay, therefore the stability and accuracy of PHIL experiments needs to be assessed before an experiment takes place. A number of compensation methods for improving the stability and the inaccuracies are frequently used within PHIL simulations \cite{Guillo2014,Viehweider2011}. Hardware components without detailed simulation models are now capable of being tested through PHIL simulations.
\end{itemize}

CHIL techniques have been successfully used for testing protection relays \cite{Almas2012}, novel control algorithms for power system devices \cite{Loddick2011}, PMU and other low power devices. However, sometimes the fidelity to represent the dynamics of complex power components such as power electronics converters have to be compromised due to the fixed time-step of real-time simulatons, limiting the size of the simulations and the transient performance \cite{Kotsampopoulos2015}. For this purpose, PHIL simulations have been carried out for testing the integration of converter connected generation into power systems or validating novel power converter structures \cite{Kotsapopoulos2012}.  
%
%
\section{Future Research Needs and Directions}
\label{sec:future_research}
Simulation techniques are essential in the engineering and validation of smart grid system development. As outlined above, different approaches are useful for different design stages. However, there are still several open issues related to the usage of simulations in the domain of power and energy systems. In the following, the most important research needs and target directions are discussed.
\subsection{Representative Models and Model Exchange}
\label{sec:representative_models}
One of the main challenges the research community faces with respect to single and multi-domain simulations is the availability of representative validated models. It is crucial that the models utilized for the purpose of simulation (offline or real-time) are validated against their representative behaviour. Most domain specific simulation tools offer elementary components required in the development of complex models. These are used by the research community to develop more complex models and novel control solutions. However, the following two challenges need to be addressed:

\begin{itemize}
	\item \textit{Interoperability:} Such validated models made available within one simulator cannot be utilized within other simulation tools. This often requires rebuilding the same model for each of the simulation tools by the prospective user.
	\item \textit{Intellectual Property (IP):} More often than not, validated models are not shared with the wider research community due to the IP issues involved. This acts as a setback to the general progress of research.
\end{itemize}

FMI provides a potential solution to both the above challenges. FMI for Model Exchange (ME), as a standard, offers a way to develop models independent of the simulation tool, where a model is converted into a C-code, compiled, supplemented with XML-schema representing the model interactions and made available for use. As most simulation tools utilize C as their lower level language, this allows for the developed models to be imported and simulated. Furthermore, it allows for models to be exported as a black-box, protecting the IP. Although FMI-ME is well accepted within the automotive industry, at this present day, many proprietary power systems simulation tools are still yet to incorporate the FMI-ME standard. Furthermore, although the models exported from simulation tool A can be run within simulation tool B (assuming simulation tool A and B support FMI-ME), it requires both simulation tool A and B to be installed on the recipient computer, limiting the capabilities of FMI-ME.

The practice of making a software open source is widely exercised within energy domains, however libraries of models are rarely open source. Open source does provide a variety of advantages that include, but are not limited to, encouraging quality and identification of prospective collaborations.
\subsection{Co-Simulation}
\label{sec:co-simulation_future}
In research, the following topics need to be addressed to meet current and upcoming requirements for co-simulation based validation of smart grid systems:

\begin{itemize}
	\item \textit{Automated Scenario Generation/Scenario Formalization:} Sets of scenarios need to be generated automatically from a relatively high-level description, if employed simulators and parameter values are known. A close-to-reality formalisation of smart grid scenarios, that can be translated into executable co-simulation scenarios (incl. validation of scenario completeness, coherence, etc.), needs to be developed.
	\item \textit{Simulation Interface/Description Standard:} A semantic description of a simulator needs to be specified. A promising interface standard is already given in form of FMI, but an additional layer providing a more direct mapping to the real world would allow automated scenario validation.
	\item \textit{Additional Research Domains:} More research needs to be carried out in terms of integration of communication systems, economic dynamics, social/psychological effects and dynamics, security-related issues, and environmental factors/dynamics into the co-simulation setup. The latter may even lead to a need of spatially resolved, GIS-based co-simulation.
	\item \textit{Performance Optimization:} The execution of different simulators in a single experiment needs to be distributed onto several machines. Parallelization of a co-simulation execution without the requirement of a single common ``data hub'' in the form of a machine that runs the scheduler needs to be realized. The design of co-simulation scenarios should allow for a formal method to select simulators with a performance as high as possible and an output accuracy as high as necessary. Surrogate modelling should facilitate the automated creation of surrogate replacements of computationally expensive simulators for performance increase.
	\item \textit{Data Management:} Data stream management should allow for analysis, aggregation and monitoring of co-simulation output during the simulation process. Through big data storage, efficient long-term storage of large data sets should be enabled, allowing for data mining operations.
	\item \textit{Visualization:} Improved tools for easy on-line monitoring and demonstration of co-simulation scenarios should be developed. Solutions for feeding co-simulation output into a control room/SCADA setup for the sake of training, testing or development should be realized.
	\item \textit{Uncertainty Quantification (UQ):} The effect of the coupling of different UQ approaches handling different simulators in a co-simulation setup should be researched. Approaches for UQ should be developed, focused on uncertainty sources related with simulator interaction (e.g. different resolutions, data transformation, instable data exchange, etc). Moreover, useful aggregation and depiction of the output of UQ studies need to be developed.
\end{itemize}
\subsection{Real-time Simulation and Hardware-in-the-Loop}
\label{sec:real-time_simulation_and_hil_future}
%
%
Offering a wide range of possibilities for validation and testing of smart grid systems, current HIL technology still has several limitations. In the European project ERIGrid\footnote{https://www.erigrid.eu}, a survey was addressed at the experts in 12 top European research institutions about current limitations and mandatory future improvements of HIL technology (PHIL and CHIL). Current open issues can be classified into four categories:

\begin{itemize}
	\item Limited capacity of HIL simulation for complex systems (computing power, complexity and synchronization) and for studies of non-linearity, high harmonics and transient phase.
	\item Limited capacity of remote HIL and geographically distributed HIL for joint experiments, mostly due to synchronization (CHIL and PHIL) and power interface stability and accuracy with respect to loop delay (PHIL).
	\item Difficulty in integration of HIL technology to the communication layer, particularly related to the synchronization of real-time and offline simulation, as well as continuous and discrete timelines.
	\item Lack of a general framework to facilitate the reusability of models, information exchange among different proprietary interfaces or among different partners of a joint HIL experiment.
\end{itemize}

Due to the aforementioned issues, HIL technology needs several aspects to be improved. Overall, the following research trends can be observed and need to be addressed:

\begin{itemize}
	\item \textit{Integration of Co-simulation and HIL:} It is expected that integrating HIL technology into co-simulation frameworks is an important contribution toward a holistic approach for experimenting with cyber-physical energy systems. Combining the strengths of both approaches, multi-domain experiments can be studied with realistic behaviours from hardware equipment under a variety of complex environments, co-simulated by appropriate and adapted simulators from the relevant domains. It will enable a complete consideration of the electrical grid to be interconnected with other domain. Until now, this is still limited. Most of the current work involving integration of HIL and co-simulation uses only a direct coupling with the real-time simulator \cite{Bian2015} or a CHIL setup \cite{Rotger2016}. 
	\item \textit{Remote and Geographically Distributed HIL:} Latency strongly influences the accuracy (HIL) and  the stability (PHIL) of a HIL test. Moreover, random packet loss due to network congestion outside of LAN may alter the information and cause malfunction at the real-time simulator, including any connected hardware \cite{Liu2017}. Up-to-now scientists have investigated the possibility of extending PHIL beyond laboratory geographical boundaries, and mostly, for latency tolerant applications (e.g., monitoring) \cite{Lundstrom2016,Palmintier2015}. These developments could be a first step in enabling the possibility of remote HIL and geograhically distributed HIL.
	\item \textit{Interoperability and Standardization:} Within a HIL-co-simulation test it is crucial to ensure seamless communication among the individual components and simulators. Additionally, when the experiments involve multiple domains or multi-laboratories, it is required to have strong interoperability between different partners \cite{Nguyen2016}. A common information model is necessary to enable seamless and meaningful communication among applications. First attempts have been made towards creating a common reference model to improve interoperability and reusability of HIL experiments \cite{Andren2014}. With these efforts towards harmonization and standardization of HIL technology, a standardized and general framework for HIL experiments can be established.
	\item \textit{Power Interface Stability and Accuracy:} Basically, the challenge here is to synchronize and compensate the loop delay, in order to stabilize the system and increase the accuracy of a test. The first step should be selection of appropriate interface algorithms and power amplification. Recommendations from \cite{de2011} should be used. Secondly, a time delay compensation method could be considered, such as introducing phase shifting, low-pass filter to the feedback signal \cite{Kotsapopoulos2012}, extrapolation prediction to compensate for time delays \cite{Ren2011}, phase advance calibration\cite{Roscoe2010} or multi-rate real-time simulation \cite{Viehweider2011}.
\end{itemize}
%
%
\section{Conclusions}
\label{sec:conclusions}
The ongoing transition towards smart grids implies significant changes in the overall energy system's architecture and infrastructure. A newly arising issue is the continuously increasing level of the system's complexity, i.e., growing number of components in various subsystems that are interrelated to each other. In order to manage, analyse and understand this novel smart grid system, simulation approaches have to be adopted. 

As outlined in the paper mainly three different simulation approaches -- multi-domain, co-simulation, and real-time hardware-in-the-loop -- are suitable tools for analysis and validating smart grids during various development steps. Despite achievements that have been made in this area, a lot issues are yet to be solved. 

The main research directions related to co-simulation can be summarized as usability improvements, standardized interfaces, performance optimization, data management, and visualization. In the domain of real-time simulation and HIL, future research should address PHIL interface improvements, remote HIL, standardization and coupling with co-simulation. In summary, there remains ample space for future research in simulation-based smart grid system validation. 
%
%
\subsubsection*{Acknowledgments.} This work is supported by the European Community’s Horizon 2020 Program (H2020/2014-2020) under project ``ERIGrid'' (Grant Agreement No. 654113). 
%
%
\bibliographystyle{splncs03}
\bibliography{literature}
\end{document}